\documentstyle[multicol,aps,psfig]{revtex}
\draft
\def\tb{\tilde{b}}
\begin{document}
\draft

\title{$Z^0$ Production as a Test of Nuclear Effects at the LHC}

\author{Xiaofei Zhang
and George Fai
}

\address{Center for Nuclear Research, Department of Physics,
Kent State University \\
Kent, Ohio 44242, USA}
\maketitle

\centerline{\date{February 1, 2002}}



\begin{abstract}
We predict the $Z^0$ transverse momentum distribution from 
proton-proton and nuclear collisions at the LHC. After demonstrating 
that higher-twist nuclear effects are very small,
we propose $Z^0$ production as a precision test 
for leading-twist pQCD in the TeV energy region.
We also point out that shadowing 
may result in unexpected phenomenology at the LHC.
\end{abstract}

\vspace{0.1in}

\begin{multicols}{2}
The physics plans for the Large Hadron Collider (LHC) at CERN, which 
is going to be the highest-energy accelerator on Earth,
include a heavy-ion program. Quantum Chromodynamics (QCD)
for both hadronic and nuclear collision will enter a new era at the LHC,
where we hope to discover new physics. However, 
``standard physics'' needs to be tested in the new, high-energy regime.

At LHC energies, perturbative QCD (pQCD) provides a powerful 
calculational tool\cite{field}. Clearly, an understanding of pQCD 
at the hadronic collision level is a prerequisite for the 
discussion of particle production in nuclear collisions.
Enhanced power corrections from 
multiple scattering in both the initial and final states,
not to mention potential new physics from the quark-gluon plasma (QGP), 
make pQCD predictions for nuclear collisions more difficult than  
for hadronic collisions.

The testing of pQCD in nuclear collisions at the LHC requires 
``clean processes'', where pQCD works well on the hadron level. 
One of the important recent advances in pQCD theory is a
reorganization of perturbative corrections (folding selected logarithmic 
contributions at all orders into exact low-order calculations), 
which is beginning to provide practical applications\cite{snowmass}.
The soft-gluon resummation for the inclusive production of colorless 
massive states\cite{CS-b,CSS-W,qz01}
may be the best understood and best tested among
these resummation techniques.
For the transverse momentum distribution of heavy bosons 
of mass $M$,
when $p_T \ll M$, the $p_T$ distribution calculated order-by-order in
$\alpha_s$ in conventional fixed-order perturbation theory
receives a large logarithm, $\ln(M^2/p_T^2)$, at every power of
$\alpha_s$, even in the leading order in $\alpha_s$.
Therefore, at sufficiently small $p_T$, the convergence of the
conventional perturbative expansion in powers of $\alpha_s$ is
impaired, and the logarithms must be resummed. 

The heavy-ion program at the LHC will make it possible to 
observe the full $p_T$ spectra of heavy vector bosons in nuclear
collisions and will provide a testing ground for resummation theory
in nuclear collisions. In the present paper, we focus on $Z^0$
production\cite{z095,vogt01}. 
Based on LHC design luminosities\cite{LHCdesign}, we estimate 
that a month of running will provide $\sim 4*10^5$ $Z^0$
events in a proton-proton ($pp$) collision, and $\sim 8*10^2$ $Z^0$ events
in a Pb+Pb collision in a $p_T$ interval of 0.5 GeV in the peak regions
of the corresponding spectra. Due to the large mass
of the $Z^0$, and no final state rescattering in its production, 
nuclear effects from final state interactions are expected to be 
small. We will show that the power corrections enhanced by initial 
state rescattering also remain small. Thus,
leading twist pQCD should work well here.
The only important nuclear effect left is the nuclear modification 
of parton distribution functions (shadowing). 
Therefore, $Z^0$ production could provide a bench mark test for pQCD
at the LHC in both $pp$ and nuclear collisions.  

Resummation of the large logarithms in QCD can be carried out either
in $p_T$-space directly\cite{Ellis-1}, or in the so-called 
``impact parameter'', 
$\tb$-space, which is a Fourier conjugate of the $p_T$-space.
Using the renormalization group
equation technique, Collins and Soper improved the $\tb$-space
resummation to resum all logarithms as singular as
$\ln^m(M^2/p_T^2)/p_T^2$ with $m \geq 0$, when
$p_T\rightarrow 0$ \cite{CS-b}. 
Collins, Soper, and Sterman (CSS) derived a formalism
for the transverse momentum distribution of vector boson production
in hadronic collisions\cite{CSS-W}. In the CSS formalism, non-perturbative
input is needed for the large $\tb$ region.  The dependence of the pQCD
results on the non-perturbative input is not weak if the original
extrapolation proposed by CSS is used. Recently, a new extrapolation
scheme was proposed, based on solving the renormalization group equation
including power corrections\cite{qz01}. Using
the new extrapolation formula, the dependence of the pQCD results
on the non-perturbative input was significantly reduced.
The results agree with Tevatron
CDF\cite{CDF-Z} and D0\cite{D0-W} data very well in the entire $p_T$
interval from $p_T \lesssim 1$ GeV to $p_T$ as large as the
the vector mass. 

For vector boson ($V$) production in a hadron collision 
$h_A + h_B$, the CSS resummation formalism yields\cite{CSS-W}:
\begin{eqnarray}
\frac{d\sigma(h_A+h_B\rightarrow V+X)}{dM^2\, dy\, dp_T^2} =
\frac{1}{(2\pi)^2}\int d^2 \tb\, e^{i\vec{p}_T\cdot \vec{\tb}}\,
\nonumber \\
\tilde{W}(\tb,M,x_A,x_B) + Y(p_T,M,x_A,x_B) \,\,\, ,
\label{css-gen}
\end{eqnarray}
where $x_A= e^y\, M/\sqrt{s}$ and $x_B= e^{-y}\, M/\sqrt{s}$, with
rapidity $y$ and collision energy $\sqrt{s}$.
In Eq.~(\ref{css-gen}), the 
$\tilde{W}$ term dominates the $p_T$ distributions
when $p_T \ll M$, and the $Y$ term gives corrections 
that are negligible
for small $p_T$, but become important when $p_T\sim M$. 

The function $\tilde{W}(\tb,M,x_A,x_B)$ can be  
calculated perturbatively for small $\tb$, but an
extrapolation to the large $\tb$ region requiring nonperturbative input
is necessary in order to complete the Fourier transform in Eq.~(\ref{css-gen}).
In oder to improve the situation, a new form was proposed\cite{qz01}
by solving the renormalization equation including power
corrections. In the new formalism, 
$\tilde{W}(\tb,M,x_A,x_B)=\tilde{W}^{pert}(\tb,M,x_A,x_B)$, when 
$\tb \leq \tb_{max}$, with 
\begin{equation}
\tilde{W}^{pert}(\tb,M,x_A,x_B) =
{\rm e}^{S(\tb,M)}\, \tilde{w}(\tb,c/\tb,x_A,x_B) \,\,\, ,
\label{css-W-sol}
\end{equation}
where all large logarithms from $\ln(1/\tb^2)$ to $\ln(M^2)$ have
been completely resummed into the exponential factor
$S(\tb,M)$, and $c$ is a constant of order
unity \cite{CSS-W}. 
\begin{eqnarray}
\tilde{W}(\tb,M,x_A,x_B)
=\tilde{W}^{pert}(\tb_{max})
F^{NP}(\tb;\tb_{max})  \,\,\, ,
\label{qz-W-sol-m}
\end{eqnarray}
where the
nonperturbative function $F^{NP}$ is given by
\begin{eqnarray}
&&F^{NP}
=\exp\bigl\{ -\ln(M^2 \tb_{max}^2/c^2) 
\left[ g_1 \left( (\tb^2)^\alpha - (\tb_{max}^2)^\alpha\right) \right.
\nonumber\\
&& \left.   +g_2 \left(\tb^2 - \tb_{max}^2\right) \right]
-\bar{g}_2 \left(\tb^2 - \tb_{max}^2\right) \bigr\}.
\label{qz-fnp-m}
\end{eqnarray}
In Eq.(\ref{qz-W-sol-m}) $\tb_{max}$ is a parameter to separate 
the perturbatively calculated part from the non-perturbative
input. Unlike in the original CSS formalism,
$\tilde{W}(\tb,M,x_A,x_B)$ is not altered 
when $\tb < \tb_{max}$, and is independent of 
the nonperturbative parameters. In addition, the $\tb$-dependence in 
Eq.~(\ref{qz-fnp-m}) is separated according to different physics
origins. The $(\tb^2)^\alpha$-dependence mimics the
summation of the perturbatively calculable leading power
contributions to the  renormalization group equations
to all orders in the running
coupling constant $\alpha_s(\mu)$. The $\tb^2$-dependence of the
$g_2$ term is a direct consequence of dynamical power corrections to the
renormalization group equations 
and has an explicit dependence on $M$. 
The $\bar{g}_2$ term represents the effect 
of the non-vanishing intrinsic parton transverse momentum.

A remarkable feature of the $\tb$-space resummation formalism is
that the resummed exponential factor $\exp[S(\tb,M)]$
suppresses the
$\tb$-integral when $\tb$ is larger than $1/M$. Therefore,
it can be shown using the saddle point method that, for a large
enough $M$, QCD perturbation theory is valid even at $p_T=0$\cite{PP-b,CSS-W}.
As discussed in Ref.s \cite{qz01,zhang-fai}, the value of the saddle point
strongly depends on the
collision energy $\sqrt{s}$, in addition to its well-known $M^2$
dependence. Because of the steep evolution of
parton distributions at small $x$, the $\sqrt{s}$ dependence of
$\tilde{W}$ in Eq.~(\ref{css-gen}) significantly decreases the value of
the saddle point and improves the
predictive power of the $\tb$-space resummation formalism at collider
energies, in particular at the LHC.

In $Z^0$ production, since final state interactions are negligible,
power correction can arise only from initial state multiple scattering.
Power corrections directly to the physical observables are proportional 
to powers of $\Lambda_{QCD}/Q$ (Q being the
physical large scale). These corrections are small for $Z^0$
production as a result of the large mass of the $Z^0$.
Power corrections to the
evolution of the renormalization group equations are proportional
to powers of $\Lambda_{QCD}/\mu$, with evolution scale $\mu$.
Therefore, physical observables carry the effect of the
latter type power corrections
for all $\mu[Q_0,Q]$, with the boundary condition at the scale $Q_0$.
Even with large mass, $Z^0$ can still carry a large effect 
of these power corrections.

Equations (\ref{qz-W-sol-m}) and (\ref{qz-fnp-m}) represent 
the most general form of $\tilde{W}$, and thus (apart from isospin and 
shadowing effects, which will be discussed later), the only way nuclear
modifications associated with scale evolution
enter the $\tilde{W}$ term is through the coefficient 
$g_2$. (Since the ${\bar g}_2$ term of Eq. (\ref{qz-fnp-m}) is 
related to the partons' intrinsic
transverse momentum, it should not have a strong nuclear dependence.) 

The parameters $g_1$ and $\alpha$ of Eq.~(\ref{qz-fnp-m})
are fixed by the requirement of continuity of the function
$\tilde{W}(\tb)$  and its derivative at $\tb=\tb_{max}$.
(The results are insensitive to changes of $\tb_{max}$ in
the interval 0.3 GeV$^{-1} \lesssim \tb_{max} \lesssim$ 0.7~GeV$^{-1}$.
We use $\tb_{max}=$ 0.5 GeV$^{-1}$.) 
The value of $g_2$ and ${\bar g}_2$ can be obtained by fitting the
low-energy Drell-Yan data.
These data can be fitted with about equal precision if the values 
${\bar g}_2=0.25\pm 0.05$ GeV$^2$ and $g_2=0.01\pm 0.005$ GeV$^2$
are taken. As the $\tb$ dependence of the  $g_2$ and ${\bar g}_2$ 
terms in Eq.~(\ref{qz-fnp-m}) is identical, it is convenient to combine
these terms and define
\begin{equation}
G_2= \ln({M^2 \tb_{max}^2\over {c^2}})g_2 + \bar{g}_2 \,\,\, .
\label{G2}
\end{equation}
Using the values of the parameters listed above, we get 
$G_2 = 0.33 \pm 0.07$ GeV$^2$ 
for $Z^0$ production in $pp$ collisions. 
The parameter $G_2$ can be considered 
the only free parameter in the non-perturbative input in Eq. (\ref{qz-fnp-m}),
arising from the power corrections in the renormalization group equations.
An impression about the importance of power corrections can be obtained by
comparing results with the above value of $G_2$ 
to those with power corrections turned off ($G_2=0$).
We therefore define the ratio
\begin{equation}
R_{G_2}(p_T) \equiv \left.
\frac{d\sigma^{(G_2)}(p_T)}{dp_T} \right/
\frac{d\sigma(p_T)}{dp_T} \,\,\, .
\label{Sigma-g2}
\end{equation}
The cross sections in the above equation and in the results presented
in this paper have been integrated over rapidity ($-2.4 \leq y \leq 2.4$) 
and invariant mass squared.
For the parton distribution functions, we use the CTEQ5M set\cite{CTEQ5} 
in the present work.

Figure 1 displays the differential cross sections 
and the corresponding $R_{G_2}$ ratio 
(with the limiting values of $G_2=0.26$ GeV$^2$
(dashed) and $G_2=0.40$  GeV$^2$ (solid))
for $Z^0$ production as functions of $p_T$ at $\sqrt s= 14$ TeV.   
The deviation of $R_{G_2}$ from unity
decreases rapidly as $p_T$ increases, and it 
is smaller than one percent for both $\sqrt{s}=5.5$~TeV (not shown)
and $\sqrt{s}=14$ TeV in $pp$ collisions, even when $p_T=0$. 
In other words, the effect of power corrections
is very small at the LHC for the whole $p_T$  region.

\begin{figure}
\begin{center}
\psfig{figure=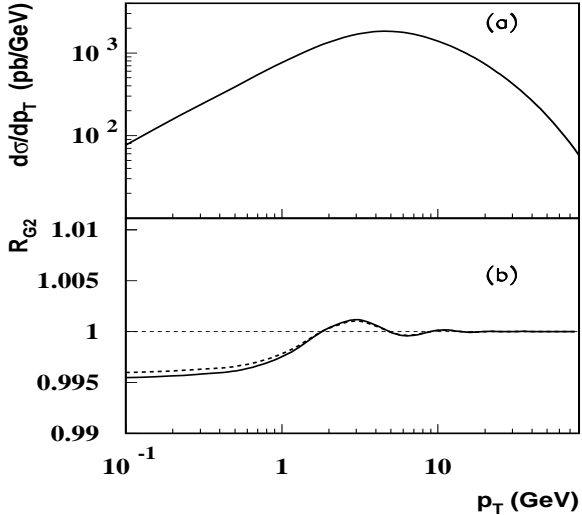,width=3.5in,height=3.0in}
\end{center}
\vspace{-0.4in}
\caption{(a) Cross section ${d\sigma / dp_T}$ for $Z^0$ production 
in $pp$ collisions at the LHC with
$\sqrt{s}=14$ TeV; (b) $R_{G_2}$ defined in Eq.~(\protect\ref{Sigma-g2})
with $G_2=$ 0.26 GeV$^2$  (dashed) and 0.40 GeV$^2$  (solid).}
\label{zfig1}
\end{figure}

In lack of nuclear effects on the hard collision,
the production of heavy vector bosons in nucleus-nucleus ($AB$) collisions
should scale, compared to the production in $pp$ collisions,
as the number of hard collisions, $AB$. 
However, there are several additional
nuclear effects on the hard collision in a
heavy-ion reaction. First of all, since
the parton distribution of neutrons is different
from that of the protons, the production cross section
of heavy vector bosons in proton-neutron interactions 
differs from the corresponding production cross section
in $pp$ collisions. This difference is the source of the
so-called isospin effects. At LHC,  $x \sim 0.02$, and
the magnitude of the isospin effects is about 2\%. This is because
when $x$ is in this range, the $u-d$ asymmetry is very small\cite{pinkbook}.

The dynamical power corrections entering the parameter $g_2$ 
should be enhanced by the nuclear
size, i.e. proportional to $A^{1/3}$. Taking into account the $A$-dependence,
we obtain $G_2 = 1.15 \pm 0.35$ GeV$^2$ for Pb+Pb reactions.
We find that with this larger value of $G_2$, 
the effects of power corrections appear to be enhanced by a factor of  
about three from $pp$ to Pb+Pb collisions at the LHC. 
Thus, even the enhanced power corrections remain
under 1\% when 3 GeV $\lesssim p_T \lesssim $ 80 GeV. This 
small effect is taken into account in the following nuclear calculations.

Next we turn to the phenomenon of shadowing, 
expected to be a function of $x$,
the scale $\mu$, and of the position in the nucleus. The latter
dependence means that in heavy-ion collisions, shadowing
should be impact parameter ($b$) dependent. The
parameterizations of shadowing in the literature take 
into account some of these effects, but no complete parameterization 
exists to date to our knowledge. For example, the HIJING 
parameterization includes impact parameter
dependence, but does not deal with the scale dependence\cite{wang1,wang2}.
On the other hand, the EKS98\cite{eks} and HKM\cite{HKM}
parameterizations have a scale 
dependence, but do not consider impact parameter dependence.
(The latter parameterizations have been compared recently\cite{eskola02}.)
In this paper we concentrate on impact-parameter integrated 
results, where the effect of the $b$-dependence of shadowing is relatively 
unimportant\cite{zfpbl02}, and we focus more attention on scale dependence. 
We therefore use EKS98 shadowing\cite{eks} in this work.

To quantify the effect of shadowing, we define
\begin{equation}
R_{sh}(p_T) \equiv \left.
\frac{d\sigma^{(sh)}(p_T,Z_A/A,Z_B/B)}{dp_T} \right/
\frac{d\sigma(p_T)}{dp_T} \,\,\, ,
\label{Sigma-sh}
\end{equation}
where $Z_A$ and $Z_B$ are the atomic numbers and $A$ and $B$ are 
the mass numbers of the colliding nuclei, and the cross section
$d\sigma(p_T,Z_A/A,Z_B/B)/dp_T$ has been averaged over $AB$,
while $d\sigma(p_T)/dp_T$ is the $pp$ cross section.
We have seen above that 
shadowing remains to be the only significant
effect responsible for nuclear modifications.  

\begin{figure}
\centerline{\psfig{figure=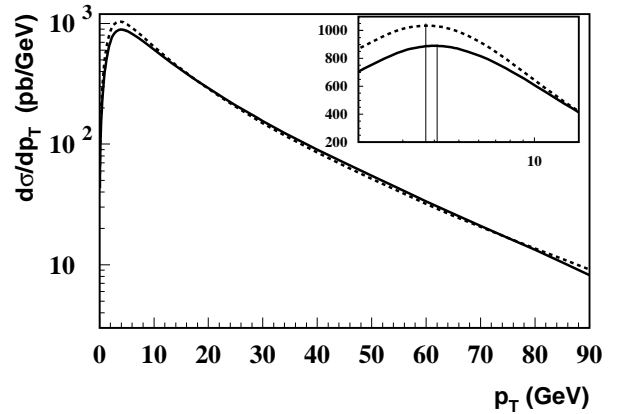,width=3.5in}}
\caption{Cross section $d\sigma^{(sh)}(p_T, Z_A/A,Z_B/B)/dp_T$
for $Z^0$ production in Pb+Pb collisions 
at the LHC with $\sqrt{s}=5.5$ TeV, averaged over $AB$ (solid line),
compared to the proton-proton cross section (dashed).}
\label{zfig12}
\end{figure} 

Figure 2 presents $ d\sigma^{(sh)}(p_T,Z_A/A, Z_B/B)/dp_T$ (solid line)
compared to  $d\sigma(p_T)/dp_T$ in $pp$ collisions (dashed) for $Z^0$ 
production at $\sqrt{s}=5.5$ TeV. The insert, which is a magnified view 
of the peak region of the cross section on a logarithmic $p_T$ scale, 
emphasizes that the shape of the distribution 
changes from $pp$ collisions. Most importantly, the peak moves form 
3.7 GeV to 4.1 GeV. 
This small shift may be difficult to detect experimentally. 
However, the peak position 
plays an important role in shadowing, due to the
steepness of the cross section. This can be seen in Fig. 3(a),
which shows the shadowing ratio (\ref{Sigma-sh}) (full line)
for $Z^0$ production at $\sqrt{s} = 5.5$ TeV. 
In Fig. 3(b) we show the $R_{G_2}$ ratio defined in Eq. (\ref{Sigma-g2})
for Pb+Pb collisions for the limiting values of $G_2=0.8$ GeV$^2$ (dashed) and
$G_2=1.5$ GeV$^2$ (solid), respectively. Since Fig. 3(b) provides a 
good measure of the overall uncertainty on the shadowing ratio,
and this uncertainty is less than 2\%,
the characteristic shape of $R_{sh}$
may be easier to confirm experimentally by comparing 
the full $p_T$ spectra in  
Pb+Pb versus $pp$ collisions at the same energy.

\begin{figure}
\begin{center}
\psfig{figure=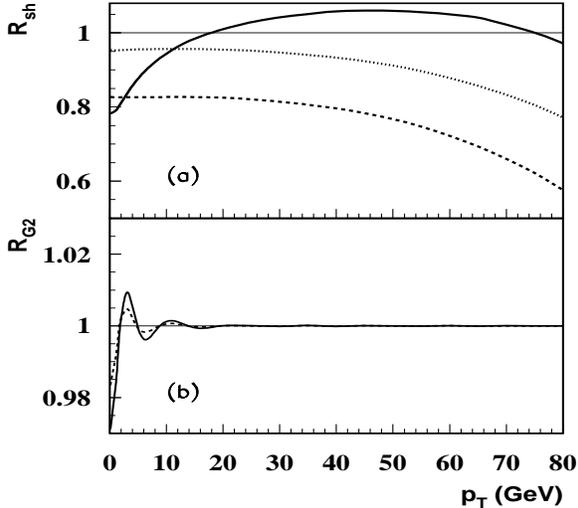,width=3.5in,height=3.0in}
\end{center}
\vspace{-0.4in}
\caption{Cross section ratios for $Z^0$ production in Pb+Pb collisions at 
$\sqrt{s}=5.5$ TeV: (a) $R_{sh}$ defined in Eq.~(\protect\ref{Sigma-sh}) 
(solid line), and $R_{sh}$ with the scale fixed at 
5 GeV (dashed) and 90 GeV (dotted);
(b) $R_{G_2}$ defined in Eq.~(\protect\ref{Sigma-g2})
with $G_2=$ 0.8 GeV$^2$ (dashed) and 1.5 GeV$^2$ (solid).}
\label{zfig3}
\end{figure}      
The appearance of $R_{sh}$ is surprising, because    
even at $p_T=90$ GeV, $x\sim 0.05$, and we are still in the ``strict
shadowing'' region. Therefore, the fact that 
$R_{sh} > 1$  for 20 GeV $\lesssim p_T \lesssim$ 70 GeV is 
not ``anti-shadowing''; rather, it is a consequence of the
change of the shape of the cross section from $pp$ to $AB$
reactions shown in Fig. 2. To better understand the shape of the ratio
as a function of $p_T$, 
we also show $R_{sh}$ with the scale fixed at the values 5 GeV (dashed 
line) and 90 GeV (dotted), respectively, in Fig. 3(a). 
In other words, the nuclear modification to the parton distribution 
function is only a function of $x$ and flavor in the calculations
represented by the dashed and dotted lines. These two curves
are similar in shape, but rather different 
from the solid line. In $\tb$ space, $\tilde{W}(\tb,M,x_A,x_B)$
is almost equally suppressed
in the whole $\tb$ region if the fixed scale shadowing is used.
However, with scale-dependent shadowing, the 
suppression increases as $\tb$ increases, as a
result of the scale $\mu\sim 1/\tb$ in the nuclear parton
distribution. We can say that the scale dependence 
re-distributes the shadowing effect. 
In the present case, the re-distribution brings $R_{sh}$ above unity 
for 20 GeV $\lesssim p_T \lesssim$ 70 GeV. When $p_T$ increases further,
the contribution from the $Y$ term starts to be important, and $R_{sh}$
dips back below one to match the fixed order pQCD result.

We see from Fig. 3 that the
shadowing effects in the $p_T$ distribution of $Z^0$ bosons
at the LHC are intimately related to the scale dependence
of the nuclear parton distributions, on which we have
only very limited data\cite{eks}. Theoretical
studies (such as EKS98) are based on the assumption 
that the nuclear parton distribution functions differ from the
parton distributions in the free proton, but obey the same DGLAP 
evolution\cite{eks}. Therefore, the tranverse momentum distribution 
of heavy bosons at the LHC in Pb+Pb collisions can provide a 
further test of our understanding of the nuclear parton distributions.

In summary, higher-twist nuclear effects appear to be 
negligible in $Z^0$ production at LHC energies. (The results
for $W^{\pm}$ production are very similar\cite{zhang-fai}.) We 
have demonstrated that the scale dependence of shadowing 
effects may lead to unexpected phenomenology of shadowing
at these energies. Overall, the $Z^0$ transverse momentum distributions
calculated in this paper can be used as a precision test 
for leading-twist pQCD in the TeV energy region for both,
proton-proton and nuclear collisions. We propose that 
measurements of $Z^0$ spectra be very high priority at the LHC. 

\vspace{0.1in}


We thank J. Qiu and R. Vogt for stimulating discussions. This work was 
supported in part by the U.S. Department of Energy under DE-FG02-86ER-40251.



\end{multicols}
\end{document}